\documentclass[12pt,twoside,fleqn]{article}
\usepackage{epsfig}
\usepackage{amstex}
\usepackage{amssymb}

\begin{document}

\begin{flushright}
BARI-TH 300/98 
\end{flushright}

\def\obs{{\cal O}}

\thispagestyle{empty}

\begin{center}
{\Large \bf {First lattice evidence for  a non-trivial\\[.6cm] renormalization 
of the Higgs condensate } }

\end{center}
\vspace{1.0cm}
\begin{center}
{\large 
P. Cea$^{1,2}$,
M. Consoli$^{3}$ and
L. Cosmai$^{2}$ \\
\vspace{1.0cm}
{\small
$^1$~Dipartimento di Fisica,  Universit\`a di Bari,  via Amendola 173, 
I 70126 Bari, Italy\\[0.05cm] 
$^2$~INFN - Sezione di Bari, via Amendola 173, I 70126 Bari, Italy \\[0.05cm]
$^3$~INFN - Sezione di Catania, Corso Italia 57, I 95129 Catania,  
Italy \\[0.05cm]} }

\end{center}
\vspace{0.8cm}
\begin{center}
{\large {\bf Abstract}}
\end{center}
\vspace{0.3cm}
General arguments related to ``triviality'' predict that, in the broken 
phase of $(\lambda\Phi^4)_4$ theory, the condensate $\langle \Phi \rangle$ 
re-scales by a factor $Z_{\varphi}$ different from the conventional 
wavefunction-renormalization factor, $Z_{\rm prop}$.  Using a lattice 
simulation in the Ising limit we measure $Z_{\varphi}=m^2 \chi$ from the 
physical mass and susceptibility and $Z_{\rm prop}$ from the residue 
of the shifted-field propagator.  We find that the two $Z$'s differ, with 
the difference increasing rapidly as the continuum limit is approached.  
Since $Z_{\varphi}$ affects the relation of $\langle \Phi \rangle$ to 
the Fermi constant it can sizeably affect the present bounds on the Higgs 
mass.  

\newpage
\setcounter{page}{1}

\section{Introduction}

Theoretical and numerical evidence 
\cite{aizen,froh,sokal,latt,luscher87,glimm,book} 
strongly supports the view that $(\lambda\Phi^4)_4$ theories are ``trivial''.  
The physical meaning of this result, however, remains controversial. 
The conventional interpretation is based on 
Renormalization-Group-Improved-Perturbation-Theory (RGIPT), while a quite 
different interpretation is advocated in Refs. \cite{zeit,primer,response}.  
The two pictures predict quite different structures for $V_{\rm eff}$, the 
effective potential of the theory.
RGIPT organizes the various contributions 
according to the perturbative/loop-expansion classification (as leading, 
next-to-leading,...terms).  Due to the lack of perturbative 
asymptotic freedom there are inescapable problems with this method when 
taking the continuum limit \cite{landau}.  The approach of Refs. 
\cite{zeit,primer,response} focuses on the class of approximations 
to $V_{\rm eff}$ that are consistent with the non-interacting nature of 
the shifted field $h(x)\equiv \Phi(x)-\langle \Phi \rangle$. 
The resulting predictions, unlike the RGIPT predictions, yield an 
excellent fit to the lattice data for $V_{\rm eff}$ 
\cite{agodi1,agodi2}. 

     A key feature of the alternative picture is the presence of ``two 
$Z$'s''.  In the {\it class} of approximations  
(including one-loop, gaussian, and post-gaussian calculations \cite{rit2}), 
where the $h$-field is governed by an effective quadratic Hamiltonian, 
the effect of bare $h\!-\!h$ self-interactions can be 
reabsorbed into the $h$-field mass $M_h$ and into the 
normalization of a physical, ``renormalized'' vacuum field \cite{V2}
\begin{equation}
\label{vR}
v_R\equiv v_B/\sqrt{Z_{\varphi}}.
\end{equation}
Here $v_B=\langle \Phi \rangle$ and $Z_{\varphi}$ is defined such that 
\begin{equation}
\label{phiR}
\left. \frac{d^2V_{\rm eff}}{d\varphi^2_R} \right|_{\varphi_R=\pm v_R}=M^2_h
\end{equation}
The role of $Z_{\varphi}$ is essential.  It provides the key-ingredient 
to get a non trivial effective potential in a ``trivial'' theory.  One 
finds \cite{zeit,primer,response,rit2,V2} that in the continuum 
limit (cutoff $\Lambda \to \infty$):
\begin{equation}
\label{Zphi}
           Z_{\varphi} \sim \ln {{\Lambda}\over{M_h}} \to \infty .
\end{equation}
Therefore, although $M^2_h/v^2_B \to 0$ in the continuum limit, in 
accord with the rigorous arguments of \cite{book}, one has
\begin{equation}
\label{vR2}
              M^2_h/v^2_R ={\rm cutoff-independent}.
\end{equation}
The divergence in Eq.(\ref{Zphi}) cannot be understood within RGIPT and 
is due to the peculiarity \cite{zeit,response} of the $p_{\mu}=0$ mode, 
responsible for spontaneous symmetry breaking. Physically, it represents 
a condensate of the symmetric-phase $\Phi$ particles \cite{zeit,gas}.  
In the $\Lambda \to \infty$ limit the infinitely increasing particle density 
in the condensate compensates for the vanishing strength of the elementary 
2-body processes (``triviality''), thus yielding a finite, negative energy 
density and a finite Higgs mass.  As such, $Z_{\varphi}$ is quite 
different from $Z_{\rm prop}$, the residue of the shifted-field propagator, 
associated with the normalization of the $p_{\mu}\neq 0$ asymptotic
one-particle states \cite{bj} and which is bounded by $Z_{\rm prop} \le 1$ 
from K\"allen-Lehmann decomposition (with ``triviality'' implying 
$Z_{\rm prop} \to 1$).  For this reason, in the presence of spontaneous symmetry
breaking, field re-scaling cannot be given as an ``operatorial'' statement
\cite{agodi1}.

    The aim of this Letter is to directly test the prediction that 
$Z_{\varphi}$ differs from $Z_{\rm prop}$.  We present the results of a 
lattice simulation of the theory (in the Ising limit) where we compute 
the mass $M_h$ and the residue $Z_{\rm prop}$ from a 2-parameter fit to 
the lattice data for the shifted-field propagator.  We then compute the 
zero-momentum susceptibility 
\begin{equation}
\label{phiB}
\left. 
{{1}\over{\chi}}=
\frac{d^2V_{\rm eff}}{d\varphi^2_B} \right|_{\varphi_B=\pm v_B}
\end{equation}
and hence obtain the dimensionless quantity
\begin{equation}
\label{m2chi}
Z_{\varphi} \equiv M^2_h \chi.  
\end{equation}
Finally, we compare $Z_{\varphi}$ with $Z_{\rm prop}$.

\section{The lattice simulation}

The one-component $(\lambda\Phi^4)_4$ 
theory   
\begin{equation}
\label{action}
   S =\sum_x \left[ {{1}\over{2}}\sum_{\mu}(\Phi(x+\hat e_{\mu}) - 
\Phi(x))^2 + 
{{r_0}\over{2}}\Phi^2(x)  +
{{\lambda_0}\over{4}} \Phi^4(x)  \right]    
\end{equation}
becomes in the Ising limit 
\begin{equation}
\label{ising}
   S_{\rm Ising} = -\kappa
\sum_x\sum_{\mu} \left[ 
\phi(x+\hat e_{\mu})\phi(x) +
\phi(x-\hat e_{\mu})\phi(x) \right]    
\end{equation}
with $\Phi(x)=\sqrt{2\kappa}\phi(x)$ and $|\phi(x)| = 1$. 

The shifted field propagator, defined at 
$p_{\mu} \neq 0$, can be computed 
as 
\begin{equation}
G(p)=\langle \sum_x \exp (ip x) h(x)h(0)\rangle
\end{equation}
for the values
$p_{\mu}={{2\pi}\over{L}}n_{\mu}$ with $n_{\mu}\neq 0 $.
An excellent fit to the lattice data is obtained (see for example Fig.1) 
by using the 2-parameter formula
\begin{equation}
G(p)= {{Z_{\rm prop}}\over{\hat{p}^2 + m^2_{\rm latt}} }
\end{equation}
where $m_{\rm latt}$ is the dimensionless lattice mass and 
$\hat{p}_{\mu}=2 \sin {{p_{\mu} }\over{2}}$.
 Finally the susceptibility $\chi$
is measured directly as
\begin{equation}
\label{suscep}
\chi_{\rm latt}=L^4 \left[ \left\langle \Phi^2 \right\rangle - 
\left\langle \Phi \right\rangle^2 \right] \
\end{equation}
with $\Phi$ the average field for each lattice configuration,
and we define
\begin{equation}
\label{Zlatt}
Z_{\varphi} \equiv m^2_{\rm latt}~ \chi_{\rm latt}
\end{equation}
To update our field configurations we used the Swendsen-Wang 
\cite{SW} cluster algorithm on $20^4$ and $24^4$ lattices.
 After discarding 10K sweeps for thermalization,
we have performed 50K sweeps, measuring our observables every 5 sweeps.
Statistical errors can be estimated through a
direct evaluation of the integrated autocorrelation time~\cite{Madras88}, 
or by using the ``blocking''~\cite{blocking} or the 
``grouped jackknife''~\cite{jackknife} algorithms.  We have checked that 
applying these three different methods we get consistent results.

We have computed at different values of the hopping parameter $\kappa$ in 
order to obtain 
a correlation length $\xi_{\rm latt}=1/m_{\rm latt}$ 
in the range 2 to $L/4$. The upper limit of
the correlation length is required in order to take under control finite-size 
effects \cite{L/4,montvay}. 
Our results
for $Z_{\varphi}$ and $Z_{\rm prop}$, in the broken phase 
$0.0751 \leq \kappa \leq 0.0764$ 
 are reported in Fig.2 and show a sizeable
difference for $m_{\rm latt} <0.3$. We have performed a consistency 
check that no such effect is 
present in the symmetric phase $0.0726\leq \kappa \leq 0.0741$, as expected. 
These other results are shown in
Fig.3. As an additional check, we have compared 
with Montvay and Weisz \cite{montvay}, 
at $\kappa=0.074$ in the symmetric phase, and got excellent agreement (see
Table 1). We also compared our results with the available data of a high statistics 
simulation in the broken phase~\cite{jansen} obtaining a rather good agreement 
(see Table 1).

\section{Conclusions}

Our numerical 
simulation of $(\lambda\Phi^4)_4$, in the Ising limit, 
shows a clear difference between two {\it measured} 
quantities: the rescaling of
the ``condensate'' $Z_{\varphi}$ 
and the more conventional quantity
$Z_{\rm prop}$ associated with the residue of the shifted field propagator.
As discussed in the introduction, this is not unexpected 
on the basis of the
alternative description of spontaneous symmetry breaking of 
Refs. \cite{zeit,primer,response}.
The effect shows up when increasing the correlation length and should become 
more and more important by approaching the continuum limit
of quantum field theory $m_{\rm latt} \to 0$.
Therefore, the relation of the lattice vacuum field $\langle\Phi\rangle$ to
the Fermi constant
and the same limits on the Higgs mass can sizeably be affected. Indeed, these
have been based on the quantity \cite{lang}
\begin{equation}
\label{rprop}
R_{\rm prop}= {{ m^2_{\rm latt}} \over{ \langle\Phi\rangle^2 }}~
Z_{\rm prop} 
\end{equation}
rather than 
\begin{equation}
\label{rphi}
R_{\varphi}= {{ m^2_{\rm latt}} \over{ \langle\Phi\rangle^2 } }~
Z_{\varphi}= {{ m^4_{\rm latt}~ \chi_{\rm latt} } \over{\langle\Phi\rangle^2}} \,.
\end{equation}

In this sense, the discovery of $Z_{\varphi}$ requires
a ``second generation'' of lattice simulations to re-check the
scaling behaviour of the various quantities 
and compare with all available theoretical descriptions of the continuum limit.

\newpage

\begin{table}
\renewcommand{\arraystretch}{2}
\begin{center}
\begin{tabular}{ccclllll}
\hline \\[-0.95cm]
\hline
Ref.                 &$L_{\text{size}}$  &$\#$ sweeps     &$\kappa$  &$m_{\text{latt}}$ &$\chi$           &$Z_\varphi$   &$Z_{\text{prop}}$ \\ \hline
Our data             &$20^4$            &$6\times10^4$   &0.074     &0.2124(60)         &142.7 $\pm$ 2.1  &0.953(55)     &0.969(6)          \\
Ref.~\cite{montvay}  &$20^3 \times 24$  &$1.6\times10^6$ &0.074     &0.2125(10)         &142.6(8)         &0.953(15)     &    -             \\[0.3cm]
Our data             &$20^4$            &$6\times10^4$   &0.076     &0.4060(61)         &38.26(74)        &0.959(34)     &0.918(6)          \\
Ref.~\cite{jansen}   &$20^4$            &$7.5\times10^6$ &0.076     &0.395(1)           &37.85(6)         &0.898(5)      &0.918(9)          \\[0.3cm]
Our data             &$20^4$            &$6\times10^4$   &0.077     &0.5718(70)         &18.20(32)        &0.916(55)     &0.899(8)          \\
Ref.~\cite{jansen}   &$16^4$            &$10^7$          &0.077     &0.563(1)           &18.18(2)         &0.887(3)      &0.886(3)          \\
\hline \\[-0.95cm]
\hline
\end{tabular}
\caption{Our lattice data compared with results available 
in the literature (Refs.\cite{montvay,jansen}). 
The values for $Z_{\text{prop}}$ in Ref.~\cite{jansen} have been evaluated by means of
renormalized perturbation theory.}
\end{center}
\label{table:I}
\end{table}
\clearpage

\section*{FIGURE CAPTIONS}

\renewcommand{\labelenumi}{Figure \arabic{enumi}.}
\begin{enumerate}
\item
The lattice data for the propagator (Eq.(9)) (circles) 
at $\kappa=0.07518$ on a $24^4$ lattice with
superimposed the fit Eq.(10) (dotted line).
\item
$Z_\varphi$ (circles) and $Z_{\text prop}$ (squares) in the broken phase 
versus $m_{\text latt}$.
\item
$Z_\varphi$ (circles) and $Z_{\text prop}$ (squares) versus $m_{\text latt}$ 
in the symmetric phase. Symbols as in Fig.2.
\end{enumerate}

\newpage
\begin{figure}[H]
\begin{center}
FIGURE 1
\end{center}
\label{Fig1}
\begin{center}
\epsfig{file=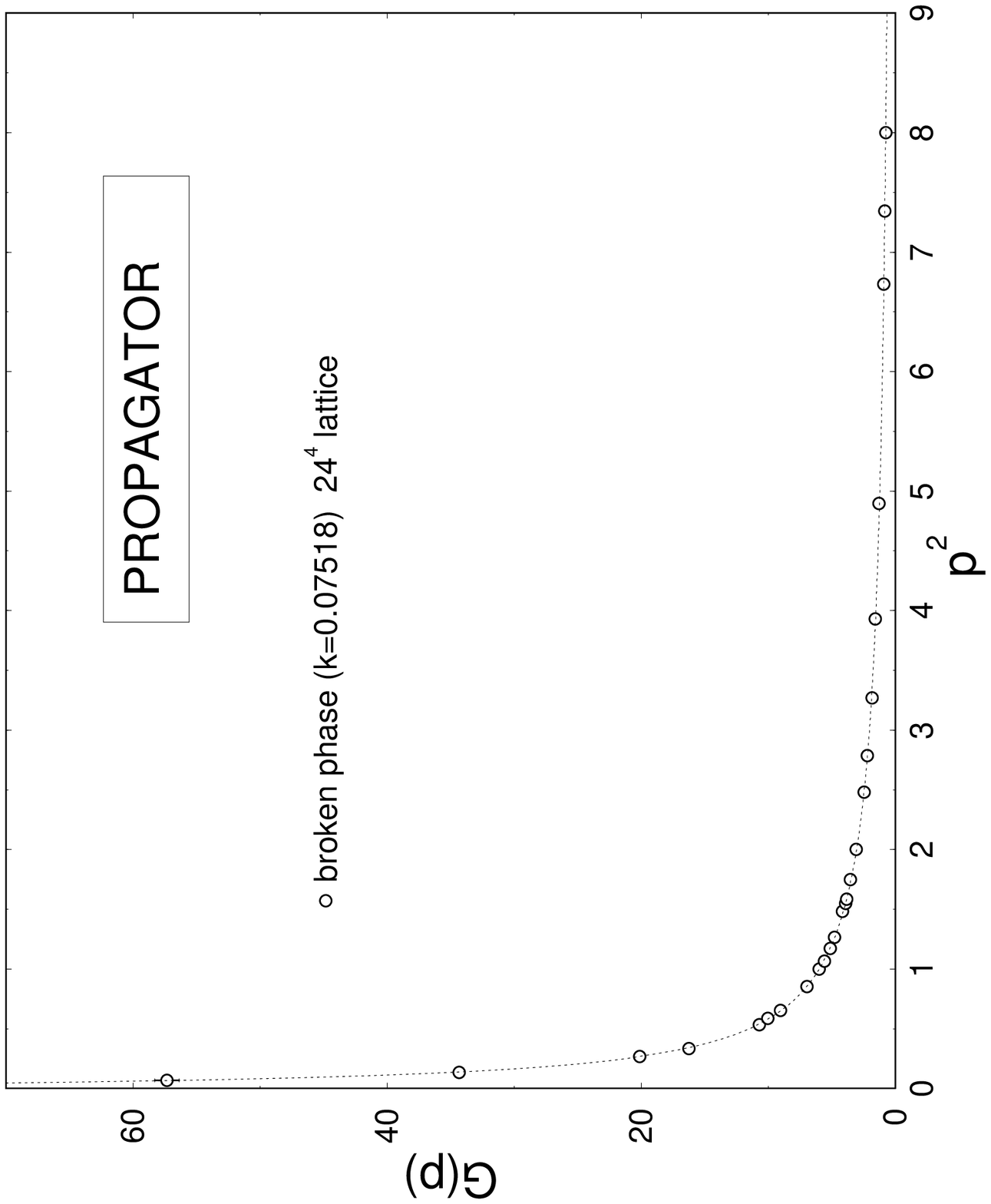,width=\textwidth}
\end{center}
\end{figure}
\newpage
\begin{figure}[t]
\begin{center}
FIGURE 2
\end{center}
\label{Fig2}
\begin{center}
\epsfig{file=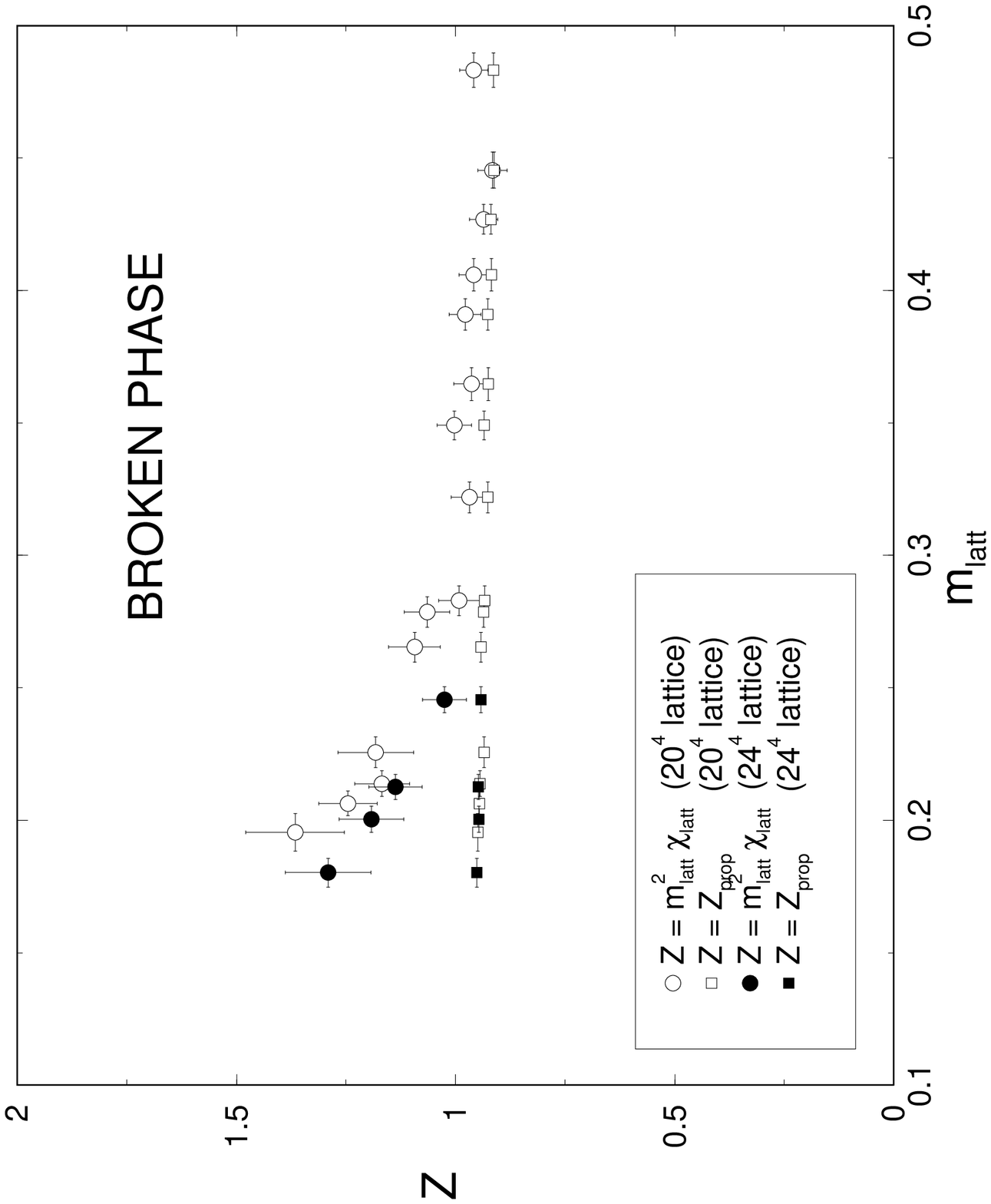,width=\textwidth}
\end{center}
\end{figure}
\begin{figure}[t]
\begin{center}
FIGURE 3
\end{center}
\label{Fig3}
\begin{center}
\epsfig{file=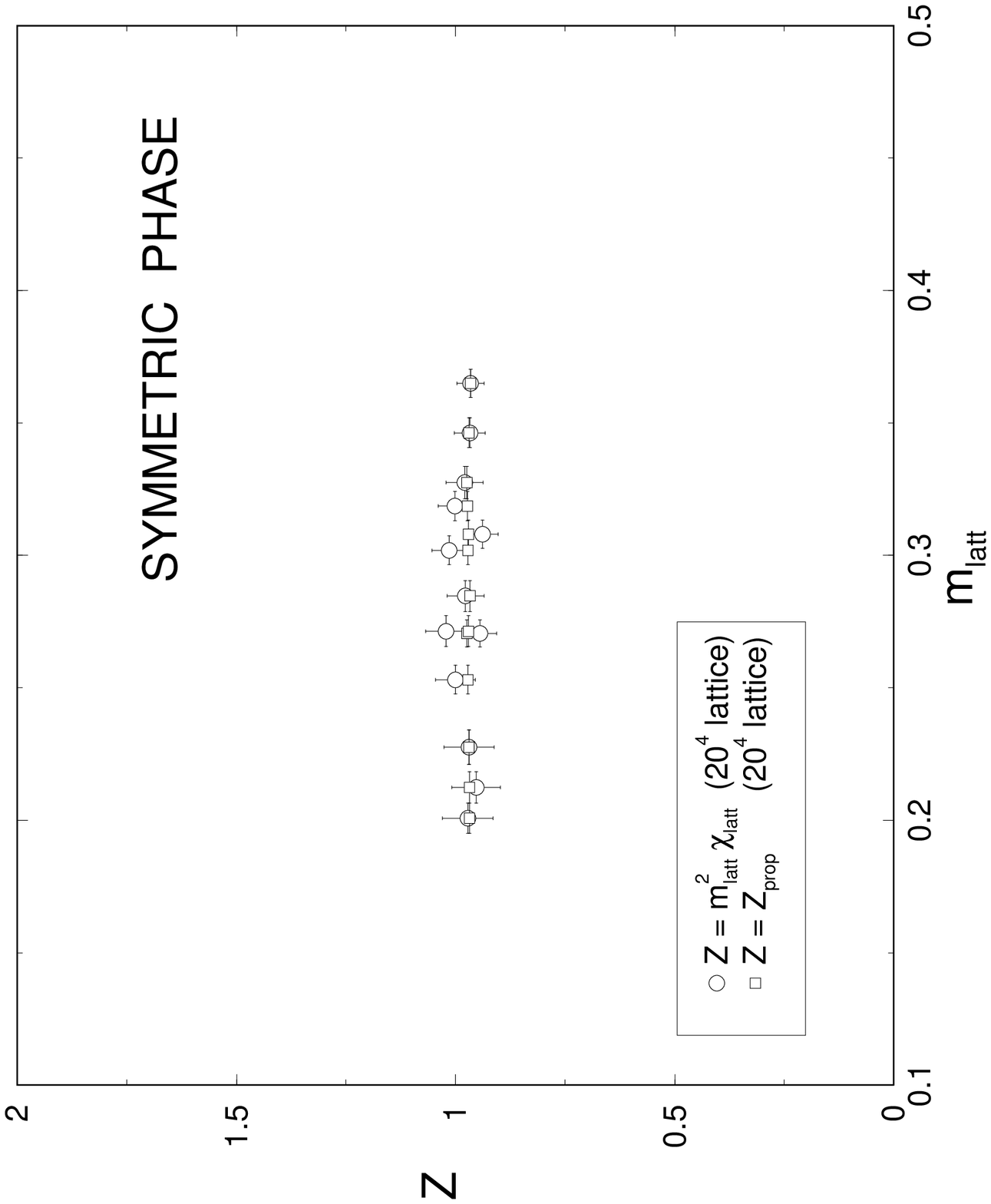,width=\textwidth}
\end{center}
\end{figure}


\vskip 40pt



\begin{thebibliography}{99}
\bibitem{aizen} M. Aizenman, Phys. Rev. Lett. {\bf 47} (1981) 1.
\bibitem{froh}  J. Fr\"ohlich, Nucl. Phys. {\bf B200}(FS4) (1982) 281.
\bibitem{sokal} A. Sokal, Ann. Inst. H. Poincar\'e, {\bf 37} (1982)
317.
\bibitem{latt}  K. G. Wilson and J. Kogut, Phys. Rep. {\bf C12} (1974)
75; G. A. Baker and J. M. Kincaid, Phys. Rev. Lett. {\bf 42} (1979)
1431; B. Freedman, P. Smolensky and D. Weingarten, Phys. Lett. {\bf
B113} (1982) 481; D. J. E. Callaway and R. Petronzio, Nucl. Phys. {\bf
B240} (1984) 577; I. A. Fox and I. G. Halliday, Phys. Lett. {\bf B
159} (1985) 148; C. B. Lang, Nucl. Phys. {\bf B 265} (1986) 630.
\bibitem{luscher87} M. L\"{u}scher and  P. Weisz, Nucl. Phys. {\bf B
290} (1987) 25; ibidem {\bf B295} (1988) 65.
\bibitem{glimm}  J. Glimm and A. Jaffe, {\em Quantum Physics: A
Functional Integral Point  of View} (Springer, New York, 1981, 2nd Ed.
1987).
\bibitem{book} R. Fern\'andez, J. Fr\"ohlich, and A. D. Sokal, {\em
Random Walks, Critical  Phenomena, and Triviality in Quantum Field
Theory} (Springer-Verlag, Berlin, 1992).
\bibitem{zeit} M. Consoli and P. M. Stevenson, Zeit. Phys. {\bf C63}
(1994) 427.
\bibitem{primer} M. Consoli and P. M. Stevenson, hep-ph/9407334.  
\bibitem{response} M. Consoli and P. M. Stevenson, 
Phys. Lett. {\bf B391} (1997) 144.
\bibitem{landau}  
M. Consoli and P. M. Stevenson, Mod. Phys. Lett. {\bf A11} (1996) 2511.
\bibitem{agodi1} A. Agodi, G. Andronico
and M. Consoli, Zeit. Phys. {\bf C66} (1995) 439.
\bibitem{agodi2} A. Agodi, G. Andronico, P. Cea, M. Consoli, L. Cosmai,
R. Fiore and P. M. Stevenson, Mod. Phys. Lett. {\bf A12} (1997) 1011. 
\bibitem{rit2} U. Ritschel, Zeit. Phys. {\bf C63} 345 (1994).
\bibitem{V2}  M. Consoli, in {\it Gauge Theories Past and Future - in 
Commemoration of the 60th Birthday of M. Veltman}, R. Akhoury et al. Eds., 
World Scientific 1992; R. Iba\~nez-Meier and P. M. Stevenson, Phys. Lett.
{\bf B297} (1992) 144; M. Consoli, Phys. Lett. {\bf B 305} (1993) 78;
V. Branchina, M. Consoli and N. M. Stivala, Zeit. Phys. {\bf C57} (1993) 251.
\bibitem{gas}
M. Consoli and P. M. Stevenson, {\it $\lambda\Phi^4$ Theory From a Particle-Gas
 Viewpoint}, hep-ph/9711449.
\bibitem{bj}
J. D. Bjorken and S. D. Drell, {\it Relativistic Quantum Fields}, McGraw-Hill 
Book Co, New York 1965, pages 132-143. 
\bibitem{SW}
R. H. Swendsen and J.-S. Wang, Phys. Rev. Lett. {\bf 58} (1987) 86.
\bibitem{Madras88} N. Madras and A. D. Sokal, J. Stat. Phys. {\bf 50} (1988) 109.
\bibitem{blocking} C. Whitmer, Phys. Rev. {\bf D29} (1984) 306;
H. Flyvbjerg and H. G. Petersen, J. Chem. Phys. {\bf 91} (1989) 461.
\bibitem{jackknife} B. Efron, {\em Jackknife, the Bootstrap and Other
Resampling Plans}, (SIAM Press, Philadelphia, 1982);
B. A. Berg and A. H. Billoire, Phys. Rev. {\bf D40} (1989) 550.
\bibitem{L/4} M. L\"uscher, Comm. Math. Phys. {\bf 104} (1986) 177.
\bibitem{montvay}
I. Montvay and P. Weisz, Nucl. Phys. {\bf B290} (1987) 327.
\bibitem{jansen}
K. Jansen, I. Montvay, G. M\"unster, T. Trappenberg and U. Wolff, Nucl. Phys. 
{\bf B322} (1989) 698.
\bibitem{lang} For a complete review see, for instance, 
C. B. Lang, {\em Computer Stochastics in Scalar Quantum
Field Theory}, in Stochastic Analysis and Application in
Physics, Proc. of the NATO ASI, Funchal, Madeira, Aug. 1993, ed. L.
Streit, Kluwer Acad. Publishers, Dordrecht 1994. 

\end{thebibliography}
\end{document}